\documentclass[11pt,dvips]{article}
\usepackage[dvips]{graphicx}
\usepackage{amsmath,amssymb,amsfonts}
\usepackage{multirow}
\usepackage{array}

\newcommand{\s}{\,{\rm s}}
\newcommand{\GeV}{\,{\rm GeV}}
\newcommand{\TeV}{\,{\rm TeV}}
\newcommand{\Mpc}{\,{\rm Mpc}}
\newcommand{\kpc}{\,{\rm kpc}}
\newcommand{\km}{\,{\rm km}}
\newcommand{\cm}{\,{\rm cm}}
\newcommand{\fex}{{\it e.g.} }
\newcommand{\ppp}[1]{\begin{pmatrix} #1 
\end{pmatrix}}

\title{\vspace{-1.5cm}
{\normalsize
\rightline{DESY 09-039}
\rightline{TUM-HEP 716/09}
}\ \vskip 1cm
\bf
Cosmic Rays from Leptophilic Dark Matter Decay via Kinetic Mixing
\vspace{11mm}} 

\author{
Alejandro~Ibarra$^{a}$, Andreas~Ringwald$^{b}$, David~Tran$^{a}$,
Christoph~Weniger$^{b}$\\[4mm]
{\normalsize\it a  Physik-Department T30d, Technische Universit\"at
M\"unchen,}\\[-0.05cm]
{\it\normalsize James-Franck-Stra\ss{}e, 85748 Garching, Germany.}\\[2mm]
{\normalsize\it b Deutsches Elektronen-Synchrotron DESY,}\\
{\it\normalsize Notkestra\ss{}e 85, 22607 Hamburg, Germany.}\\[2mm]
\date{\empty}
}

\begin{document}
\begin{titlepage} 
  \maketitle
\begin{abstract}
  If interpreted in terms of decaying dark matter, the steep rise in the
  positron fraction of cosmic rays above $10\GeV$, as observed by the PAMELA
  experiment, suggests an underlying production mechanism that favors leptonic
  channels. We consider a scenario where a portion of the dark matter is made
  of the gauginos of an unbroken hidden-sector $U(1)_X$, which interact with
  the visible sector only through a tiny kinetic mixing. The second component
  of the dark matter is made of neutralinos, and depending on the mass
  spectrum, the lightest neutralino or the hidden gaugino becomes unstable and
  subject to decay. We analyze the cosmic rays, namely the contributions to
  the positron, the extragalactic gamma-ray and the antiproton flux, which
  potentially result from these decays and demonstrate that the production of
  antiprotons can be naturally suppressed. Furthermore, we briefly discuss the
  apparent double-peak structure of the ATIC data in light of cascade-decaying
  hidden gauginos, as well as possible signatures at Fermi.
\end{abstract}

\thispagestyle{empty}
\end{titlepage}
\newpage \setcounter{page}{2}

\section{Introduction}
The PAMELA collaboration has recently reported a measurement of the positron
fraction~\cite{A08} which shows a significant excess at high energies compared
to the expectations from spallation of cosmic rays on the interstellar
medium~\cite{MS98}. This result confirms the excess in the positron fraction
reported in the past by several experiments: HEAT 94/95/00~\cite{B04},
CAPRICE94~\cite{BCF+00} and AMS-01~\cite{A07}. More importantly, the PAMELA
collaboration has provided an accurate measurement of the energy spectrum of
the positron fraction, revealing a steep rise between 7 and 100\GeV, possibly
extending to even higher energies. 

The excess in the positron fraction has been interpreted as an indication for
the annihilation~\cite{annihilation1, BE99, HMS+06, annihilation2} or the
decay~\cite{IT08a, decay, ADD+08} of dark matter particles. If interpreted as
dark matter annihilations, the dark matter particle must annihilate into $W$,
$Z$, $e$, $\mu$, $\tau$ for any dark matter mass, or into $q$, $b$, $t$, $h$
for a multi-TeV dark matter mass.  In both cases one needs large boost
factors, \fex from dark matter substructures, since the required annihilation
rates are much larger than the cross-section suggested by cosmology,
$\langle\sigma v\rangle\approx 3\times 10^{-26} {\rm cm}^3/\s$~\cite{CKR+09},
would lead one to expect.  On the other hand, if interpreted as dark matter
decay, the dark matter particles must have a mass larger than $\sim300\GeV$, a
lifetime around $10^{26}$s and the decay must proceed preferentially into
charged leptons of the first and second generation~\cite{IT08b}. The
properties of the dark matter particles are also subject to constraints from
the diffuse gamma-ray flux as measured by the EGRET instrument~\cite{SMR04},
and the antiproton flux, as measured by PAMELA~\cite{A09}, BESS95~\cite{M98},
BESS95/97~\cite{O00}, CAPRICE94~\cite{BCF+97}, CAPRICE98~\cite{BBS+01} and
IMAX~\cite{M96}. More concretely, the good agreement of the theoretical
predictions for the antiproton flux with the measurements suggests that the
dark matter particle annihilates or decays mostly into leptons.  Moreover,
taking additionally into account the excess in the total electron+positron
flux as observed by the ATIC experiment~\cite{C08a} hints to dark matter
particle masses above $1\TeV$.\footnote{For the combined electron+positron
flux, we use results from measurements that have been undertaken by ATIC,
HEAT94/95~\cite{D01}, PPB-BETS~\cite{T08}, H.E.S.S.~\cite{A08a},
BETS~\cite{T01}, CAPRICE94~\cite{BCF+00} and SANRIKU~\cite{K99}.} Although the
interpretation of the positron excess in terms of dark matter is very
suggestive, it should be borne in mind that nearby astrophysical sources such
as pulsars might produce sizable positron fluxes in the energy range explored
by PAMELA~\cite{pulsars}.

In this paper we will discuss the cosmic-ray signatures of supersymmetric
scenarios with a gaugino of an unbroken hidden-sector $U(1)$ gauge group which
interacts with the visible sector only via a tiny kinetic mixing~\cite{IRW08}.
We will speculate that the tiny kinetic mixing induces the decay of the dark
matter particles into lighter supersymmetric particles and positrons, thus
providing a potential explanation to the excess observed by PAMELA.  Assuming
exact $R$-parity conservation, two possibilities may arise.  First, we will
study the case that the hidden gaugino mass is smaller than the lightest
neutralino mass. We will show that the neutralino mostly decays into two
charged leptons and the hidden gaugino.  The hard positrons produced in the
decay can then potentially explain the steep rise in the positron fraction
observed by PAMELA and the absence of an excess in the antiproton flux.
Secondly, we will study the case that the lightest neutralino mass is smaller
than the hidden gaugino mass. If this is the case, it is the hidden gaugino
which decays into the lightest neutralino, either directly or in a cascade
decay, when there are supersymmetric particles with masses between the hidden
gaugino mass and the lightest neutralino mass.  Either particle could be the
dominant component of dark matter. However, we will focus in this work on the
possibility that the dominant component of dark matter is the lightest
neutralino, which may allow direct dark matter detection. 

In section 2 we will shortly review the main features and possible origins of
scenarios with a hidden abelian gauge group and kinetic mixing. In section 3
we will review the procedure to calculate the gamma-ray, positron and
antiproton fluxes at Earth from dark matter decay. In section 4 we will
present our results for the decaying neutralino case and the decaying hidden
gaugino case. Lastly, in section 5 we will present our conclusions.

\section{Model}
Many extensions of the Minimal Supersymmetric Standard Model (MSSM)
contemplate the possibility of a hidden sector, consisting of superfields
which are singlets under the Standard Model gauge group. Hidden sector
superfields usually couple very weakly to our observable sector, thus
constituting a very natural arena for finding dark matter candidates. We
consider an extension of the MSSM by a hidden abelian gauge group $U(1)_X$
(for details about the model see Ref.~\cite{IRW08}). This gauge group remains
unbroken at low energies and couples to the MSSM only through a tiny kinetic
mixing, $\chi$, with the hypercharge $U(1)_Y$~\cite{Holdom86}:
\begin{eqnarray}
  SU(3)_c\times SU(2)\times 
  \underbrace{U(1)_Y \times U(1)_X}_\text{kin. mixing $\chi$}.
  \label{eqn:GaugeGroup}
\end{eqnarray}
We assume that all matter states charged under $U(1)_X$ are heavy and
negligible. As a consequence the $U(1)_X$ gauge boson completely decouples
from the observable sector. However, as shown in Ref.~\cite{IRW08}, a non-zero
mass-mixing of the order of $\delta M\sim\mathcal{O}(\chi\cdot M_X)$ between
hidden gaugino and bino generally remains.  More precisely, in the basis where
the kinetic terms are canonical, the extended $(5\times5)$ neutralino mass
matrix reads, to lowest order in $\chi$, 
\begin{eqnarray}
  \mathcal{M}_\text{N}=\ppp{ M_X & \delta M & 0 & 0 & 0 \\ 
  \delta M & M_1 & 0 & -M_Z c_\beta s_W & M_Z s_\beta s_W \\
  0 & 0 & M_2 & M_Z c_\beta c_W & -M_Z s_\beta c_W \\
  0 & -M_Z c_\beta s_W & M_Z c_\beta c_W & 0 & -\mu \\
  0 & M_Z s_\beta s_W & -M_Z s_\beta c_W & -\mu & 0 }\;.
  \label{eqn:NeutralinoMassMatrix}
\end{eqnarray}
Here, $\mu$ denotes the MSSM $\mu$-term, $M_Z$ the mass of the $Z^0$ gauge
boson, $M_1$ and $M_2$ the bino and wino masses, respectively, $s_W$ the sine
of the Weinberg angle and $s_\beta$ is related to the ratio of the two Higgs
VEVs.

Below, we will concentrate on the case where the lightest supersymmetric
particle in the visible sector is a bino-like neutralino $\chi_1^0$. Then,
depending on the masses of the hidden gaugino and of the neutralino, one of
the two particles becomes unstable with a lifetime that is roughly given by
\begin{equation}
  \tau_{X,\chi^0_1} \sim \mathcal{O}(10^{-2} - 10)\times 10^{26}\s  
  \cdot\left(
  \frac{M_{X,\chi^0_1}}{100\GeV} \right)^{-1} \left(
  \frac{\theta}{10^{-24}} \right)^{-2} \;,
  \label{eqn:lifetime}
\end{equation}
where we made use of the mixing angle $\theta\simeq\delta
M/|M_{\chi^0_1}-M_X|\sim\mathcal{O}(\chi)$. The exact prefactor depends on the
dominant decay modes and on the mass spectrum of the supersymmetric particles.
However, it is apparent that a lifetime around $10^{26}\s$, as required to fit
the PAMELA excess with decaying dark matter particles, implies an extremely
small mixing of the order of $\chi\sim10^{-24}$. Note that the mixing must be
somewhat larger when the decaying particle is only a subdominant component of
the dark matter, but an upper bound of roughly $\sim10^{-20}$ holds from the
requirement that the particle has not already decayed.

The thermal production of hidden gauginos by oscillations between bino and
hidden gaugino, which generally takes place in the primeval MSSM plasma, is
irrelevant for the mixing parameters that we are looking at~\cite{IRW08}.
However, the hidden gaugino may be produced non-thermally, \fex in the decay
of a heavy gravitino, or it may be a thermal relic of the hidden
sector~\cite{FTY08}. In the latter case one requires additional particles that
are charged under the hidden $U(1)_X$ with masses around $M_X$, however. For
simplicity, we will assume throughout this work that the lightest neutralino
in the visible sector has the right energy density to make up the dominant
part of the observed dark matter, whereas the abundance of the hidden gaugino
is always subdominant, $\rho_X\ll \rho_{\chi_1^0} \simeq \rho_\text{DM}$.

The actual mass scale of the hidden gaugino depends on how the breaking of
supersymmetry is mediated to the visible sector and the hidden $U(1)_X$. If
the soft masses $\sim M_\text{soft}$ in the visible sector arise from gauge
mediation, whereas $U(1)_X$ couples to the supersymmetry-breaking sector only
gravitationally, the predicted mass hierarchy is $M_\text{soft}\gg m_{3/2}
\sim M_X$, where $m_{3/2}$ denotes the mass of the gravitino. In this scenario
the lightest neutralino cannot be dark matter, because it would decay into the
gravitino in the early Universe.  However, if we suppose gravity mediation to
the visible sector and anomaly mediation~\cite{RS99} to the $U(1)_X$, one
expects $M_\text{soft}\sim m_{3/2}\gg M_X$. If the soft masses of both the
visible sector and the $U(1)_X$ arise from gravity mediation, one in general
expects that all masses are of the same order $M_\text{soft}\sim m_{3/2} \sim
M_X$. Below we will assume that the gravitino is heavy enough to have no
impact on the decay modes, $m_{3/2}>\max(M_X,M_{\chi^0_1})$. \\

Additional hidden sector $U(1)$ gauge factors are a generic feature of string
compactifications. For example, in the ``mini-landscape'' of orbifold
compactifications of the heterotic string~\cite{L07} one encounters, at the
compactification scale, a breaking of the gauge symmetry to a theory involving
many hidden $U(1)$s, \fex ${\rm E}_8\times {\rm E}_8\to {\rm G}_{\rm SM}
\times {\rm U(1)}^4 \times [{\rm SO(8)}\times {\rm SU(2)}\times {\rm U(1)}^3]$
and the like. Similarly, type II compactifications generically invoke hidden
sector $U(1)$s, often also for global consistency requirements. Some of these
hidden $U(1)$s may remain unbroken down to very small scales~\cite{AJK+08}. 

Kinetic mixing is generated by the exchange of heavy messengers that couple
both to the hypercharge $U(1)_Y$ as well as to the hidden $U(1)_X$.
Correspondingly, it is loop suppressed, $\chi = g_Y g_X\,C/(16\pi^2)$, where
$g_Y$ and $g_X$ are the abelian gauge couplings and $C$ is a dimensionless
constant. In field-theoretic setups, the latter is naturally of order
one~\cite{Holdom86} and thus way too large for our purposes. However, it can
be much smaller if there are additional gauge or global symmetries (cf.
Refs.~\cite{IRW08,CTY08a}). Moreover, in models arising from string
compactifications rather small mixings seem to be generic~\cite{DKM97, AS04,
AJK+08, AGJ+08}. In the context of compactifications of the heterotic string,
the size of the mixing has been estimated as~\cite{DKM97}
\begin{equation}
  \chi \sim \frac{g_Y g_X}{16\pi^2} \frac{\Delta m}{M_P}\,,
\end{equation}
where $\Delta m\ll M_P$ is the mass splitting in the messenger sector and
$M_P$ the Planck scale. Indeed, the mixing is quite small in this case (\fex
$\chi\sim 10^{-16}$, if $\Delta m$ is associated with gauge mediated
supersymmetry breaking, $\Delta m\sim 100$~TeV), but still to large for our
purposes. However, a sufficiently strong suppression of the coupling could be
achieved in models with multiple hidden $U(1)$s~\cite{IRW08}.

In compactifications of type-II string theories, where the hidden $U(1)$s
arise via D-branes in the (extra-dimensional) bulk that have no intersection
with the branes responsible for the visible sector, one has to distinguish two
cases.

{\em i)} In large volume scenarios, where the hidden D3-branes and the branes
supporting hypercharge are located at generic positions in some Calabi-Yau,
kinetic mixing is suppressed by the large volume $V_6$ of the extra
dimensions~\cite{AGJ+08},
\begin{equation}
  \chi \simeq \frac{g_Y g_X}{16\pi^2} \left( V_6 M_s^6 \right)^{-2/3}\,,
  \label{generic_LARGE_volume}
\end{equation}
where $M_s$ is the string scale, which is related to the Planck scale via
$M_s\sim g_s M_P/\sqrt{V_6 M_s^6}$, where $g_s$ is the string coupling. We
thus see that we may obtain a kinetic mixing of order $10^{-24}$, if we assume
$V_6 M_s^6 \sim 10^{32}$, corresponding to a string scale of order $M_s \sim
$~TeV. However, this case is problematic because the mass of the gravitino is
of the order of meV~\cite{CQS05}, with impact on the stability of
supersymmetric dark matter candidates like the neutralino or hidden gaugino.

{\em ii)} In scenarios with significant warping, such as KKLT~\cite{KKL+03},
the standard model stack of branes, notably the brane featuring the
hypercharge $U(1)$, is placed at a special position --  at the tip of a warped
throat -- while the hidden brane is separated from it by a distance $d$ along
the throat. In this case kinetic mixing may be exponentially
suppressed~\cite{AGJ+08} like
\begin{equation}
  \chi \sim \frac{g_Y g_X}{16 \pi^2}e^{-md}\;,
  \label{warped_throat} 
\end{equation} 
reminiscent of a ``Yukawa type" interaction.  In fact, in warped
compactifications, the closed string fields that mediate the kinetic mixing,
notably the NS-NS $B$ form and the R-R $C$ form fields,  acquire masses, 
\begin{equation}
  m \sim n\,M_s/(V_3 M_s^3),
\end{equation} 
for some integer $n$, from the vacuum expectation values of their three-form
field strengths, $H_3 = dB_2$, $F_3=dC_2$, which are threading three-cycles of
volume $V_3$ and cause the warping. Therefore, we conclude that no strong
fine-tuning is required to obtain a tiny kinetic mixing in a KKLT-like
scenario.  Indeed, the estimate~(\ref{warped_throat}) yields the desired small
value $\chi\sim 10^{-24}$ for~$m d\sim 48$.\\

As mentioned above, depending on the mass spectrum of the supersymmetric
particles, either the lightest neutralino, $\chi_1^0$, or the hidden gaugino,
$X$, becomes unstable. The relevant decay modes are shown in
Tab.~\ref{tab:DecayChains}. These decays may be detectable as anomalous
contributions to the cosmic-ray fluxes observed at Earth. We will discuss this
in some detail in the next section.

\begin{table}[h]
  \centering
  \begin{tabular}{b{.0cm}b{4.0cm}|b{4.0cm}b{.0cm}}
    \hline
    &\multicolumn{2}{c}{\bf Neutralino / Hidden gaugino decay
    modes}\vspace{+0.0cm}&\\ 
    &
    \centering
    $M_{\chi^0_1}>M_X$ \vspace{-0.5cm} & 
    \centering
    $M_{\chi^0_1}<M_X$ \vspace{-0.5cm}& \\&
    \begin{eqnarray}
      \nonumber
      \chi_1^0\rightarrow\left\lbrace
      \begin{aligned}
        f&\tilde{\bar{f}}^{\ast}_{L/R} \rightarrow f\bar{f}X\\
        X&h^0\\
        X&Z^0
      \end{aligned}
      \right.
    \end{eqnarray}
    \vspace{0.2cm}
    &
    \begin{eqnarray}
      \nonumber
      X\rightarrow\left\lbrace
      \begin{aligned}
        f&\tilde{\bar{f}}^{(\ast)}_{L/R}\rightarrow f\bar{f} \chi_i^0\\
        \chi^0_i&h^0\\
        \chi^0_i&Z^0\\
        \chi^\pm_j&W^\mp
      \end{aligned}
      \right.
    \end{eqnarray}&\\
  \end{tabular}
  \caption{Dominant decay modes. Depending on the masses of the hidden
  gaugino, $M_X$, and the lightest neutralino, $M_{\chi_1^0}$, one of the two
  particles becomes unstable with a lifetime roughly given by
  Eq.~\eqref{eqn:lifetime}. Since the three-body decay into fermion pairs
  $f\bar{f}$ is mainly mediated by virtual sfermions, $\tilde{f}^\ast$, we
  show this explicitly. Furthermore, when a sfermion is lighter than the
  decaying particle, the corresponding three-body decay crosses over to a
  cascade decay. The subsequent decay and fragmentation of the Higgs- and
  gauge bosons, charginos and neutralinos is not shown. Note that the letter
  $f$ represents any lepton or quark.}
  \label{tab:DecayChains}
\end{table}

\section{Propagation Models}
The decay of dark matter particles induced by the kinetic mixing discussed
above will generate potentially sizable numbers of Standard Model particles
which may be detectable in cosmic-ray experiments if the dark matter decay
rate is high enough.  In the following, we will consider the cosmic-ray
signatures resulting from the decay of neutralinos and hidden gauginos in
three different channels, namely gamma rays, positrons and
antiprotons\footnote{Dark matter decay also produces a flux of neutrinos which
are unfortunately unobservable due to the large atmospheric neutrino
backgrounds~\cite{CGI+08}.}.  Interestingly, in all of these channels,
experiments are now reaching sensitivities to dark matter lifetimes of the
order of $10^{26}~\text{s}$.

\paragraph{Gamma rays.}
Photons from the decay of GeV -- TeV mass dark matter particles will manifest
themselves as an anomalous contribution to the diffuse extragalactic gamma-ray
background.  More specifically, there are two distinct contributions to the
diffuse gamma-ray flux from dark matter decay. There is an anisotropic
component from the decay of dark matter particles in the Milky Way halo, as
well as an isotropic diffuse emission from dark matter decaying at
cosmological distances.  The latter contribution is red-shifted due to the
expansion of the Universe.  Thus, for a dark matter mass $m_{\text{DM}}$ and a
dark matter lifetime of $\tau_{\text{DM}}$, the gamma-ray flux will be given
by the sum \cite{gamma-ray}
\begin{equation}
  \left[E^2 \frac{dJ}{dE}\right]_{\text{DM}} = \left[E^2
  \frac{dJ}{dE}\right]_{\text{halo}} + \left[E^2
  \frac{dJ}{dE}\right]_{\text{extra}}\;,
\end{equation}
where the halo component is given by
\begin{equation}
  \left[E^2 \frac{dJ}{dE}\right]_{\text{halo}} = \frac{2 E^2}{m_{\text{DM}}}
  \frac{dN_\gamma}{dE} \frac{1}{8 \pi \tau_{\text{DM}}} \int_{\text{los}}
  \rho_{\text{DM}}(\vec{l}) d\vec{l}\;,
\end{equation}
in which $\rho_{\text{DM}}$ is the dark matter halo profile, which in the
following we assume to be an Navarro-Frenk-White (NFW) \cite{NFW96} profile
for definiteness. The results will have only a slight dependence on the
particular choice of halo profile. $dN_\gamma/dE$ denotes the energy spectrum
of photons from dark matter decay.  The integration extends along the line of
sight (los), yielding an angular dependence of the signal. For our results, we
will average the anisotropic halo contribution over the whole sky excluding a
band of $\pm 10^\circ$ around the Galactic disk. The extragalactic component
is given by
\begin{equation}
  \left[E^2 \frac{dJ}{dE}\right]_{\text{extra}} = \frac{2 E^2}{m_{\text{DM}}}
  C_\gamma \int_1^\infty dy \frac{dN_\gamma}{d(yE)} \frac{y^{-3/2}}{\sqrt{1 +
  \Omega_\Lambda / \Omega_{\text M} y^{-3}}}
\end{equation}
with the coefficient
\begin{equation}
  C_\gamma = \frac{\Omega_{\text{DM}} \rho_{\text c}}{8 \pi \tau_{\text{DM}}
  H_0 \Omega_{\text M}^{1/2}} \;.
\end{equation}
The integration is over the red-shift $z$, with $y = z + 1$.  In these
expressions, $\Omega_\Lambda\simeq 3\Omega_\text{M}$, $\Omega_{\text{M}}\simeq
0.24$ and $\Omega_{\text{DM}}\simeq0.20$ are the vacuum, matter and dark
matter density parameters, respectively, while $H_0=70.1 \km \s^{-1}
\Mpc^{-1}$ is the present value of the Hubble parameter and $\rho_{\text c}$
is the critical density. One finds that the halo component is dominant,
although both contributions are of the same order.  For the background to the
dark matter signal from conventional astrophysical sources, we will assume a
power law spectrum that fits the data points below 1 GeV,
\begin{equation}
  \left[E^2 \frac{dJ}{dE}\right]_{\text{background}} = 6.8 \times 10^{-7}
  E^{-0.32} ~\text{GeV}~(\text{cm}^2~\text{str}~\text{s})^{-1}\;.
\end{equation}

\paragraph{Antimatter propagation.}
At the energies of interest here, only antiparticles created within the Milky
Way's dark matter halo will be of importance.  Antimatter propagation in the
Galaxy is a fairly complicated process due to diffusion, energy loss and
annihilation effects. It is commonly described using a stationary two-zone
diffusion model with cylindrical boundary conditions~\cite{GDB+}. Under this
approximation, the number density of antiparticles per unit kinetic energy,
$f(T,\vec{r},t)$, satisfies the following transport equation, which applies
both for positrons and antiprotons:
\begin{eqnarray}
  \label{transport}
  \frac{\partial f}{\partial t}&=&
  \nabla \cdot [K(T,\vec{r})\nabla f] + \\\nonumber &&+
  \frac{\partial}{\partial T} [b(T,\vec{r}) f]
  -\nabla \cdot [\vec{V_c}(\vec{r})  f]
  -2 h \delta(z) \Gamma_{\rm ann} f+Q(T,\vec{r})\;.
\end{eqnarray}
We assume free escape conditions for the solution $f(T,\vec{r},t)$ at the
boundary of the diffusion zone, which is approximated by a cylinder with
half-height $L = 1-15~\rm{kpc}$ and radius $ R = 20 ~\rm{kpc}$, and solve the
equation for the steady-state case, where $\partial f/ \partial t=0$.

The first term on the right-hand side of the transport equation is the
diffusion term, which accounts for the propagation through the tangled
Galactic magnetic field.  The diffusion coefficient $K(T,\vec{r})$ is assumed
to be constant throughout the diffusion zone and is parameterized by:
\begin{equation}
  K(T)=K_0 \;\beta\; {\cal R}^\delta\;,
\end{equation}
where $\beta=v/c$ is the velocity in units of the speed of light $c$, and
${\cal R}$ is the rigidity of the particle, which is defined as the momentum
in GeV per unit charge, ${\cal R}\equiv p[{\rm GeV}]/Z$.  The normalization
$K_0$ and the spectral index $\delta$ of the diffusion coefficient are related
to the properties of the interstellar medium and can be determined from flux
measurements of other cosmic-ray species, mainly from the Boron-to-Carbon
(B/C) ratio~\cite{MDT+01}. The second term accounts for energy losses due to
inverse Compton scattering on starlight or the cosmic microwave background,
synchrotron radiation and ionization.  The third term is the convection term,
which accounts for the drift of charged particles away from the disk induced
by the Milky Way's Galactic wind.  It has axial direction and is also assumed
to be constant inside the diffusion region: $\vec{V}_c(\vec{r})=V_c\; {\rm
sign}(z)\; \vec{k}$.  The fourth term accounts for antimatter annihilation
with rate $\Gamma_{\rm ann}$, when it interacts with ordinary matter in the
Galactic disk, which is assumed to be an ``infinitely'' thin disk with
half-height $h=100$ pc.  Lastly, $Q(T,\vec{r})$ is the source term of
positrons or antiprotons which is given by
\begin{equation}
  Q(E,\vec{r})=\frac{\rho_{\text{DM}}(\vec{r})}{m_{\text{DM}}
  \tau_{\text{DM}}}\frac{dN}{dE}\;,
  \label{source-term}
\end{equation}
where $dN/dE$ is the energy spectrum of antiparticles created via the decay of
dark matter particles.  In the transport equation, reacceleration effects and
non-annihilating interactions of antimatter in the Galactic disk have been
neglected.

The solution of the transport equation at the Solar System, $r=r_\odot$,
$z=0$, can be formally expressed by the convolution
\begin{equation}
  f(T)=\frac{1}{m_{\text{DM}} \tau_{\text{DM}}}
  \int_0^{T{\rm max}}dT^\prime G(T,T^\prime) 
  \frac{dN(T^\prime)}{dT^\prime}\;,
  \label{solution}
\end{equation}
where $T_{\rm max}=m_{\text{DM}}$ for the case of the positrons and $T_{\rm
max}=m_{\text{DM}}-m_p$ for the antiprotons.  The solution is thus factorized
into two parts.  The first part, given by the Green's function
$G(T,T^\prime)$, encodes all of the information about the astrophysics (such
as the details of the halo profile and the propagation of antiparticles in the
Galaxy).  The remaining part depends exclusively on the nature and properties
of the decaying dark matter particles, namely the mass, the lifetime and the
energy spectrum of antiparticles produced in the decay.

We will consider the dark matter lifetime to be a free parameter that is
constrained by requiring a qualitatively good agreement of the predicted
positron fraction with the PAMELA results.  Therefore, the only uncertainties
in the computation of the antimatter fluxes stem from the determination of the
Green's function, \fex from the uncertainties in the propagation parameters
and the halo profile. As it turns out, the uncertainties in the precise shape
of the halo profile are not crucial for the determination of the primary
antimatter fluxes, since the Earth receives mostly antimatter created within a
few kpc from the Sun, where the different halo profiles are very similar. On
the other hand, the uncertainties in the propagation parameters can
substantially change the predictions for the antimatter fluxes, even by as
much as two orders of magnitude for the antiproton flux.  The reason for this
large uncertainty is a correlation among the diffusion parameters and the size
of the diffusion zone. The ranges of the astrophysical parameters that are
consistent with the B/C ratio and that produce the maximal, median and minimal
positron and antiproton fluxes are listed in 
Tab.~\ref{tab:param-positron}~and~\ref{tab:param-antiproton}~\cite{DLD+08}.

Positrons and antiprotons have different properties regarding their
propagation, and their respective transport equations can be approximated by
different limits of Eq.~\eqref{transport}. By exploiting the cylindrical
symmetry of the problem, it is then possible to find semi-analytical solutions
to the transport equation in each case. Approximate interpolating functions
for the Green's function can be found in Ref.~\cite{IT08a}.

\paragraph{Positron Flux.}

\begin{table}[h]
  \begin{center}
    \begin{tabular}{|c|ccc|}
      \hline
      Model & $\delta$ & $K_0\,({\rm kpc}^2/{\rm Myr})$ & $L\,({\rm kpc})$ \\
      \hline 
      M2 & 0.55 & 0.00595 & 1  \\
      MED & 0.70 & 0.0112 & 4 \\
      M1 & 0.46 & 0.0765 & 15  \\
      \hline
    \end{tabular}
    \caption{\label{tab:param-positron}\small 
    Astrophysical parameters compatible with the B/C ratio that yield 
    the minimum (M2), median (MED) and maximal (M1) flux of positrons.}
  \end{center}
\end{table}

For the case of the positrons, Galactic convection and annihilations in the
disk can be neglected in the transport equation, which is then simplified to:
\begin{equation}
  \nabla \cdot [K(T,\vec{r})\nabla f_{e^+}] +
  \frac{\partial}{\partial T} [b(T,\vec{r}) f_{e^+}]+Q(T,\vec{r})=0\;,
  \label{transport-positron}
\end{equation}
where the rate of energy loss, $b(T,\vec{r})$, is assumed to be a spatially
constant function parameterized by $b(T)=\frac{T^2}{T_0\tau_E}$, with
$T_0=1\;{\rm GeV}$ and $\tau_E=10^{16}\;{\rm s}$.

The solution to this equation is formally given by the convolution
Eq.~\eqref{solution}. The explicit form of the Green's function
is~\cite{HMS+06}
\begin{equation}
  G_{e^+}(T,T^\prime)=\sum_{n,m=1}^\infty 
  B_{nm}(T,T^\prime) 
  J_0\left(\zeta_n \frac{r_\odot}{R}\right) 
  \sin\left(\frac{m \pi}{2 }\right),
  \label{greens-function}
\end{equation}
where $J_0$ is the zeroth-order Bessel function of the first kind, whose
successive zeros are denoted by $\zeta_n$. On the other hand,
\begin{eqnarray}
  B_{nm}(T,T^\prime)&=&\frac{\tau_E T_0}{T^2}
  C_{nm} \times \\\nonumber
  &&\times\exp\left\{\left(\frac{\zeta_n^2}{R^2} + \frac{m^2 \pi^2}{4
  L^2}\right) 
  \frac{K_0 \tau_E}{\delta - 1} 
  \left[\left(\frac{T}{T_0}\right)^{\delta-1}
  -\left(\frac{{T^\prime}}{T_0}\right)^{\delta-1}\right]\right\},
\end{eqnarray}
with
\begin{equation}
  C_{nm}=\frac{2}{J_1^2(\zeta_n)R^2 L} \int_0^R r^\prime dr^\prime 
  \int_{-L}^L dz^\prime  \rho(\vec{r}\,^\prime)
  J_0\left(\zeta_n \frac{r^\prime}{R}\right) 
  \sin\left[\frac{m \pi}{2 L}(L-z^\prime )\right]\;,
\end{equation}
where $J_1$ is the first-order Bessel function.

The interstellar positron flux from dark matter decay is finally given by
\begin{equation}
  \Phi_{e^+}^{\rm{prim}}(T) = \frac{c}{4 \pi m_{\text{DM}} \tau_{\text{DM}}} 
  \int_0^{m_{\text{DM}}}dT^\prime G_{e^+}(T,T^\prime) 
  \frac{dN_{e^+}(T^\prime)}{dT^\prime}\;.
\end{equation} 
The dependence of the positron flux on the diffusion model is mostly important
at low energies, where the signal lies well below the background.

Rather than measuring the positron flux, most experiments measure the positron
fraction, $\Phi_{e^+}/(\Phi_{e^-}+\Phi_{e^+})$, since most sources of
systematic errors, such as detector acceptance or trigger efficiency, cancel
out when computing the ratio of particle fluxes. The background to the
positron flux from dark matter decay will be constituted by a secondary
positron flux originating from the collision of primary protons and other
nuclei on the interstellar medium.

For the background fluxes of primary and secondary electrons, as well as
secondary positrons, we use the parameterizations obtained in Ref.~\cite{BE99}
from the GALPROP numerical  code for cosmic-ray propagation~\cite{MS98}. We
leave the normalization of the primary electron flux as a free parameter to be
fitted in order to match the observations of the positron fraction.

\begin{table}[h]
  \begin{center}
    \begin{tabular}{|c|cccc|}
      \hline
      Model & $\delta$ & $K_0\,({\rm kpc}^2/{\rm Myr})$ & $L\,({\rm kpc})$
      & $V_c\,({\rm km}/{\rm s})$ \\
      \hline 
      MIN & 0.85 & 0.0016 & 1 & 13.5 \\
      MED & 0.70 & 0.0112 & 4 & 12 \\
      MAX & 0.46 & 0.0765 & 15 & 5 \\
      \hline
    \end{tabular}
    \caption{\label{tab:param-antiproton} \small Astrophysical parameters
    compatible with the B/C ratio that yield the minimal (MIN), median (MED)
    and maximal (MAX) flux of antiprotons.}
  \end{center}
\end{table}

\paragraph{Antiproton flux.}
The general transport equation, Eq.~\eqref{transport}, can be simplified by
taking into account that energy losses are negligible for antiprotons.
Therefore, the transport equation for the antiproton density, $f_{\bar
p}(T,\vec{r},t)$, is then given by:
\begin{equation}
  0=\frac{\partial f_{\bar p}}{\partial t}=
  \nabla \cdot (K(T,\vec{r})\nabla f_{\bar p})
  -\nabla \cdot (\vec{V_c}(\vec{r})  f_{\bar p})
  -2 h \delta(z) \Gamma_{\rm ann} f_{\bar p}+Q(T,\vec{r})\;,
  \label{transport-antip}
\end{equation}
where the annihilation rate, $\Gamma_{\rm ann}$, is determined by the Galactic
Hydrogen and Helium densities and the proton-antiproton scattering
cross-section, for which we use the parameterization by Tan and
Ng~\cite{TN83}.

Analogously to the positron case, the solution to the transport equation can
be expressed as a convolution of the form Eq.~\eqref{solution}. The analytic
expression for the Green's function reads~\cite{D01a}:
\begin{eqnarray}
  G_{\bar p}(T,T^\prime)&=&\sum_{i=1}^{\infty}
  {\rm exp}\left(-\frac{V_c L}{2 K(T)}\right)\times\\\nonumber
  &&\times
  \frac{y_i(T)}{A_i(T) {\rm sinh}(S_i(T) L/2)} 
  J_0\left(\zeta_i \frac{r_{\odot}}{R}\right)\delta(T-T^\prime)\;,
\end{eqnarray}
where
\begin{eqnarray}
  y_i(T)&=&\frac{4}{J^2_1(\zeta_i)R^2}
  \int_0^R r^\prime \,dr^\prime\; J_0\left(\zeta_i \frac{r^\prime}{R}\right)
  \times\\
  \nonumber
  &&\times\int_0^L dz^\prime {\rm exp}
  \left(\frac{V_c (L-z^\prime)}{2 K(T)}\right)
  {\rm sinh}\left(\frac{S_i(L-z^\prime)}{2}\right)
  \rho(\vec{r}\,^\prime)\;,
\end{eqnarray}
and
\begin{eqnarray}
  A_i(T)&=&2 h \Gamma_{\rm ann}(T) + V_c+k S_i(T) 
  {\rm coth} \frac{S_i(T) L}{2}\;,\\
  S_i(T)&=&\sqrt{\frac{V_c^2}{K(T)^2}+\frac{4 \zeta_i^2}{R^2}}\;.
\end{eqnarray}
The interstellar antiproton flux is then given by
\begin{equation}
  \Phi_{\bar p}^{\rm IS}(T)=\frac{v_{\bar p}(T)}{4\pi m_{\text{DM}}
  \tau_{\text{DM}}} 
  \int_0^{m_{\text{DM}}-m_p}dT^\prime G_{\bar p}(T,T^\prime) 
  \frac{dN_{\bar p}(T^\prime)}{dT^\prime}\;,
\end{equation}
where $v_{\bar p}(T)$ is the antiproton velocity.  However, in order to
compare the calculated antiproton spectrum with experimental results, one also
has to take into account the effect of solar modulation.  In the force field
approximation~\cite{GA67,GA68} the effect of solar modulation can be included
by applying the following relation between the antiproton flux at the top of
the Earth's atmosphere and the interstellar antiproton flux~\cite{Perko87}:
\begin{equation}
  \Phi_{\bar p}^{\rm TOA}(T_{\rm TOA})=
  \left(
  \frac{2 m_p T_{\rm TOA}+T_{\rm TOA}^2}{2 m_p T_{\rm IS}+T_{\rm IS}^2}
  \right)
  \Phi_{\bar p}^{\rm IS}(T_{\rm IS}),
\end{equation}
where $T_{\rm IS}=T_{\rm TOA}+\phi_F$, with $T_{\rm IS}$ and $T_{\rm TOA}$
being the antiproton kinetic energies at the heliospheric boundary and at the
top of the Earth's atmosphere, respectively, and $\phi_F$ being the solar
modulation parameter, which varies between 500 MV and 1.3 GV over the
eleven-year solar cycle. Since experiments are usually undertaken near solar
minimum activity, we will choose $\phi_F=500$ MV for our numerical analysis in
order to compare our predicted flux with the collected data.  For the
antiprotons, we will not examine the background from spallation, but simply
require that the fluxes from dark matter decay lie appropriately below the
measurements so as to be compatible with predominantly secondary antiproton
production.

\section{Results}
Below, we will present our results for the cosmic-ray signatures of decaying
neutralinos and decaying hidden gauginos. In both cases, we will start with an
analysis of the predictions that follow when assuming that the visible sector
is described by an exemplary point in the coannihilation region of the mSUGRA
parameter space. This ensures a consistent cosmology in the visible sector and
that all free parameters of the MSSM are fixed. After that, we will go beyond
this mSUGRA scenario and discuss how the cosmic-ray signatures can change in
more generic cases. This will include a discussion about cascade decay in
light of the apparent double-peak structure of the ATIC data.

\subsection{Decaying Neutralinos}
The potentially relevant decay modes for the case $M_X < M_{\chi^0_1}$, where
the lightest neutralino can decay into the hidden gaugino, are summarized in
Tab.~\ref{tab:DecayChains}. Beside three body decays, which produce
fermion/anti-fermion pairs, we also have to take into account the decay into
Higgs- and $Z^0$ bosons.  Throughout the analysis we will assume that the
lightest neutralino $\chi_1^0$ makes up the dominant part of the dark matter,
$\rho_{\chi_1^0}\simeq \rho_\text{DM}$.

\paragraph{mSUGRA point.}
As stated above, our exemplary mSUGRA model lies in the coannihilation region.
The defining parameters are $m_0 = 150\GeV$, $m_{1/2} = 720 \GeV$, $A_0 = 0$,
$\tan\beta=10$ and sign$\,\mu=+1$. In this model the lightest neutralino has a
mass of $301\GeV$ and the correct relic density to be dark matter, $\Omega h^2
\simeq 0.104$.\footnote{The mass spectrum and relic abundance were calculated
with the aid of DarkSUSY~5.0.4~\cite{G04}.} As typical for models in the
coannihilation region, the three right-handed sleptons have masses around $304
- 307 \GeV$, which is similar to the mass of the lightest neutralino.
Furthermore, this particular mSUGRA point, which features \fex a
spin-independent cross section per proton of $2.6\times 10^{-46}\cm^2$, can be
probed with the next-to-next generation direct DM detection experiments like
XENON1T and LUX/ZEP.

\begin{table}[h]
  \centering
  \begin{tabular}{ccccccc}
    \hline
    \multirow{2}{*}{$M_X$ [GeV]} & \multicolumn{5}{c}{Branching Ratios for
    $\chi_1^0 \rightarrow$}& \multirow{2}{*}{$\tau_{\chi_1^0}\;[10^{26}\s]$} \\
    & $e^-e^+X$ 
    & $\mu^-\mu^+X$ 
    & $\tau^-\tau^+X$ 
    & $h^0 X$ 
    & $Z^0 X$ 
    \\\hline
    1   & 28\%& 28\%& 32\%& 8.8\%& 2.6\% &1.8\\
    50  & 27\%& 27\%& 30\%&  13\%& 2.4\% &1.7\\
    100 & 24\%& 24\%& 28\%&  21\%& 2.4\% &1.5\\
    150 & 21\%& 21\%& 24\%&  32\%& 2.6\% &1.3\\
    200 & 30\%& 30\%& 36\%&  --- & 3.7\% &\\
    \hline
  \end{tabular}
  \caption{Branching ratios for the decay of a neutralino $\chi_1^0$ into a
  lighter hidden gaugino $X$, for different hidden gaugino masses $M_X$. In
  the visible sector, masses and mixing parameters are fixed by a mSUGRA
  scenario in the coannihilation region as described in the text. The lightest
  neutralino has a mass of $301\GeV$. Branching ratios of three-body decays
  into neutrinos, $\chi_1^0\rightarrow\nu\bar{\nu}X$, and quarks,
  $\chi_1^0\rightarrow q\bar{q}X$, are smaller than $0.3\%$ and $0.02\%$,
  respectively. The two-body decay into photons, $\chi_1^0 \rightarrow \gamma
  X$, is one-loop suppressed and neglected. We also indicate the lifetime of
  the neutralino which gives the best fit to the data.}
  \label{tab:DNmSUGRA}
\end{table}

The dominant branching ratios of the neutralino decay are summarized in
Tab.~\ref{tab:DNmSUGRA}, for different masses of the hidden
gaugino.\footnote{The calculations were done with FeynArts~3.4 and
FormCalc~5.4~\cite{Hahn01, HP99}.} Most interestingly, the fraction of decays
into charged leptonic final states is never below $\sim 65\%$. Beside the
small masses of the right-handed sleptons, the underlying reason is the large
$\mu$-term, $\mu=865\GeV$, which suppresses the mixing between the bino-like
lightest neutralino $\chi_1^0$, the hidden gaugino and the higgsinos like
$\sim\mathcal{O}(M_Z/\mu)$.

To obtain the energy distribution of gamma rays, positrons and antiprotons
that are produced in the neutralino decay, we use the event generator PYTHIA
6.4~\cite{SMS06}. From these spectra, the contribution to cosmic-ray fluxes as
measurable at Earth can be derived as described in the previous section. Note
that the lifetime of the neutralino is always fixed by requiring a
qualitatively good agreement with the positron fraction as measured by PAMELA.

\begin{figure}[h]
  \centering
    \includegraphics[width=\linewidth]{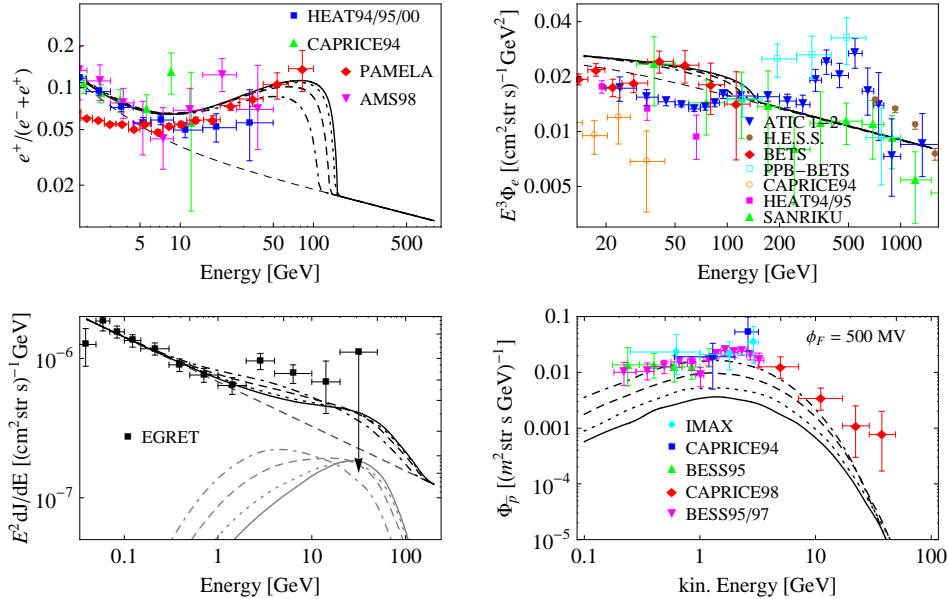}
  \caption{Positron fraction, total electron+positron, extragalactic gamma-ray
  and antiproton flux of a decaying neutralino $\chi_1^0$ as predicted for our
  exemplary mSUGRA scenario. The used branching ratios are shown in
  Tab.~\ref{tab:DNmSUGRA}. The mass of the decaying neutralino is $301\GeV$,
  the hidden gaugino mass varies between $1\GeV$ (solid), $50\GeV$ (dotted),
  $100\GeV$ (dashed) and $150\GeV$ (dot-dashed). We used the MED propagation 
  model. In the lower left plot, the grey lines indicate the flux without
  background. In the lower right plot, we only show the flux without
  background.}
  \label{fig:DNmSUGRA}
\end{figure}

Our results are shown in Fig.~\ref{fig:DNmSUGRA}. We find that in principle
the model can account for the observed excess in the positron fraction around
$10-100\GeV$ if the hidden gaugino is light with a mass $M_X\lesssim 50\GeV$,
although the predicted peak seems to rise too slowly to fully match the PAMELA
data. This slow rise is due to the two-body decay into Higgs bosons, whose
subsequent fragmentation produces rather soft positrons.  From the lower plots
of Fig.~\ref{fig:DNmSUGRA} it is apparent that the model is compatible with
the EGRET measurements of the extragalactic gamma-ray
background.\footnote{Gamma rays with energies below $\sim10\GeV$ stem from the
fragmentation of the Higgs boson whereas gamma rays at higher energies mainly
come from $\tau$ decay.} However, the contribution to the antiproton flux can
be problematic for hidden gaugino masses above $\sim 100\GeV$.\footnote{ Note
that the uncertainty in the antiproton flux at Earth from dark matter decay
can be as large as one order of magnitude in both directions~\cite{IT08a}, due
to our ignorance of the precise propagation parameters.} Of course, the peak
in the ATIC data around $300-800\GeV$ cannot be reproduced in this setup.

\paragraph{Idealized three-body decay of a heavy neutralino.}
For different parameters of the underlying MSSM model, the above plots can
mainly change in two ways. Firstly, a larger value of the $\mu$-parameter
would reduce the branching ratio into Higgs- and $Z^0$ bosons.\footnote{ A
concrete lower bound on the $\mu$-parameter is extremely model-dependent.
However, in the concrete scenario with a bino-like lightest neutralino where
we take $\tan\beta=10$, $\alpha_h=-0.1$ and assume that the right-handed
sleptons have a mass around $1.02\cdot M_{\chi^0_1}$, the lower bound
$\mu\gtrsim \mathcal{O}(2M_{\chi^0_1})$ turns out to be sufficient to suppress
the branching ratio into Higgs-bosons below $20\%$.} As a result, the rise in
the positron fraction would be steeper, and the contribution to the antiproton
flux smaller. Secondly, a higher mass of the decaying neutralino would shift
the peak to higher energies, as suggested by the ATIC data.

\begin{figure}[h]
  \centering
  \includegraphics[width=\linewidth]{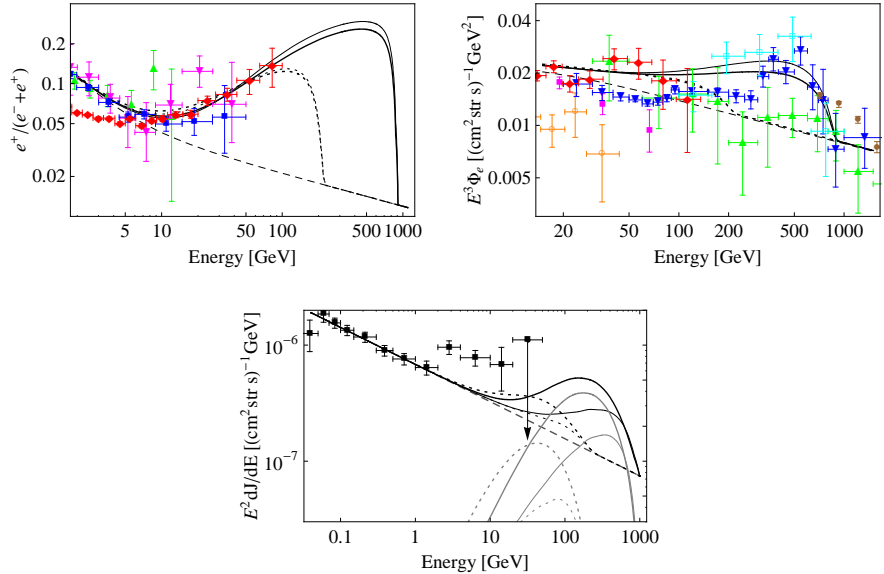}
  \caption{Positron fraction, total electron+positron flux and extragalactic
  gamma-ray flux for an idealized, three-body decaying bino-like neutralino.
  We neglect effects from $h^0$ and $Z^0$ bosons and assume pure democratic
  three-body decay into charged lepton pairs. The masses of the neutralino and
  the hidden gaugino are $500\GeV$ and $150\GeV$ (thick solid lines) or
  $1850\GeV$ and $300\GeV$ (thick dotted lines), respectively. The thin lines
  show the predictions when the decay into the tau-channel is neglected. The
  mass of the right-handed sleptons is assumed to be by a factor $1.1$ larger
  than the neutralino mass.}
  \label{fig:DNheavy}
\end{figure}

In Fig.~\ref{fig:DNheavy} we show our results for the cosmic-ray fluxes in the
idealized case where a bino-like neutralino decays only via virtual
right-handed sleptons. This resembles scenarios with a large $\mu$ term and
large masses for the left-handed sleptons. The masses of the hidden gaugino
and the neutralino $\chi_1^0$ are $150\GeV$ and $500\GeV$ (solid lines) or
$300\GeV$ and $1850\GeV$ (dotted lines), respectively. Note that the thick
lines correspond to the standard case where the neutralino decays
democratically into all three flavors.

As expected, the rise in the positron fraction is now steeper and can easily
accommodate the PAMELA data. Furthermore, a very heavy neutralino around
$1.8\TeV$ allows to also account for the ATIC excess.  In any case we find a
clear excess in the extragalactic gamma-ray flux at energies above $10\GeV$.

The gamma rays come mainly from $\tau$ decays and bremsstrahlung, but the
latter is a subdominant effect as long as the three-body decay into charged
leptons is democratic. However, the decay into taus can be suppressed in cases
where the stau mixing angle is large, since the correspondingly larger
left-handed component of the lighter stau weakens the coupling to the
bino-like neutralino and the hidden gaugino. For example, if the lighter stau
is equally left- and right-handed, the three-body decay into taus would be
suppressed by a factor of $\sim0.5$. To obtain a lower bound on the predicted
gamma-ray signal, we also show the case where the neutralino decays into the
first two generations only (see thin lines in Fig.~\ref{fig:DNheavy}). The
gamma-ray flux is much smaller in this case and comes mainly from
bremsstrahlung of the electrons produced in the three-body decay
$\chi_1^0\rightarrow e^- e^+ X$. Note that in any case we obtain a tight
correlation between the contributions to the positron flux and the
extragalactic gamma-ray flux, where the latter comes mainly from $\tau$ decays
in most cases.\\

Heavy bino-like neutralinos with masses above a few hundred GeV are
problematic for cosmology, since they are typically overproduced, even when
coannihilation with sleptons is taken into account. At the same time, wino-
and higgsino-like lightest neutralinos do not exhibit the desired leptophilic
decay.\footnote{Winos only couple to left-handed sleptons, which are typically
heavier than the right-handed ones, whereas higgsinos can easily decay into
the Higgs boson.} However, these problems are absent if one considers
scenarios where the hidden gaugino is heavier than the lightest neutralino,
$M_X>M_{\chi^0_1}$. First, due to the mixing with the bino, the interactions
of the hidden gaugino are automatically ``bino-like''. Second, for the small
mixings that we consider bounds from overproduction arguments are
irrelevant~\cite{IRW08}. Note that the results from this paragraph can also
hold in that case, provided one exchanges the roles of the hidden gaugino and
the lightest neutralino.  However, this requires that all sparticles, apart
form the lightest neutralino, are heavier than the hidden gaugino.
Generically this will not be the case and the hidden gaugino will
cascade-decay through the different sparticles down into the lightest
neutralino. We will consider this in detail in the next subsection.

\subsection{Decaying Hidden Gauginos}
A hidden gaugino that is heavier than the lightest neutralino,
$M_X>M_{\chi^0_1}$, turns out to be more appealing from the phenomenological
point of view. In this case, the mass of the lightest neutralino can be small
and of the order of a few $100\GeV$, and the hidden gaugino automatically
possesses the ``bino-like'' interactions which are desirable for the
leptophilic decay.  We will again assume that the lightest neutralino makes up
most of the dark matter, whereas the hidden gaugino contributes only a
subdominant part $\rho_X\ll\rho_{\chi_1^0}$ to the overall matter density of
the Universe.  Note that in this case the lifetime of the hidden gaugino can
be as small as $\tau_X\sim 10^{17}\s$, the current age of the Universe,
provided that its relic abundance is small enough. For definiteness, we will
take $\rho_X=10^{-3}\rho_{\chi_1^0}$ throughout this section. 

We will firstly discuss the contributions from the decaying hidden gaugino to
the cosmic-ray flux as predicted for our reference mSUGRA scenario. Secondly
we will consider the multi-peak structure of a cascade-decaying hidden gaugino
in light of the ATIC data.

\paragraph{mSUGRA point.} 
The considerations in this paragraph are again based on the mSUGRA scenario
described above. Depending on the mass of the hidden gaugino, its decay can
produce fermions, neutralinos, charginos, Higgs- and gauge bosons as depicted
in Tab.~\ref{tab:DecayChains}. The corresponding branching ratios are
summarized in Tab.~\ref{tab:DHmSUGRA}, where we do not show the subsequent
decays of the neutralinos $\chi_{2,3,4}^0$ and charginos $\chi^\pm_{1,2}$ for
simplicity.\footnote{These subsequent decays are taken into account in our
calculations. We singled out the dominant decay modes in our reference mSUGRA
model and used them in the PYTHIA code: $X\rightarrow h^0 \chi_4^0$,
$X\rightarrow Z^0 \chi_3^0$, $X\rightarrow W^\mp \chi^\pm_2$,
$\chi_3^0\rightarrow\chi^0_1 Z^0$, $\chi_4^0\rightarrow\chi^0_1 h^0$ and
$\chi_2^\pm \rightarrow \chi_1^\pm Z^0 (28\%),\; \chi_1^\pm h^0 (27\%),\;
\chi_2^0 W^\pm (36\%)$. The decay of $\chi_2^0$ and $\chi^\pm_1$ only produces
leptons and is neglected.} 

\begin{table}[h]
  \centering
  \begin{tabular}{cccccccc}
    \hline
    \multirow{2}{*}{$M_X$ [GeV]} & \multicolumn{6}{c}{Branching Ratios for
    $\tilde{X}\rightarrow$} & \multirow{2}{*}{$\tau_X$\;[$10^{23}\s$]} \\
    & $\nu\tilde{\nu}$ 
    & $l\tilde{l}$ 
    & $q\tilde{q}$ 
    & $h^0\chi^0_i$ 
    & $Z^0\chi^0_i$ 
    & $W^\pm\chi^\mp_i$ 
    \\\hline
    600  &   1.8\%& 98.2\%& ---  &  0.1\%& 0.0\%&   ---  & 1.1\\ 
    700  &   5.6\%& 92.9\%& ---  &  0.6\%& 0.0\%&   0.9\%& \\
    800  &   5.6\%& 84.6\%& ---  &  3.5\%& 0.2\%&   6.1\%& 1.0\\
    850  &   0.7\%& 49.8\%& ---  & 17.3\%& 1.2\%&  31.0\%& \\
    900  &  15.3\%& 53.7\%& ---  & 10.7\%& 0.9\%&  19.4\%& 0.8\\
    1000 &  14.1\%& 81.1\%& ---  &  1.4\%& 1.0\%&   2.4\%& \\
    1200 &  13.3\%& 76.8\%& ---  &  2.7\%& 2.5\%&   4.7\%& 0.7\\
    1400 &  13.2\%& 74.1\%& 1.6\%&  2.9\%& 2.8\%&   5.4\%& \\
    1600 &  12.5\%& 68.5\%& 8.4\%&  2.7\%& 2.7\%&   5.2\%& \\
    \hline
  \end{tabular}
  \caption{Branching ratios of the dominant decay modes of a hidden gaugino
  that is cascade-decaying into the MSSM particle zoo. The underlying scenario
  is our chosen mSUGRA reference point as described in the text. Neutrinos and
  charged leptons decay essentially democratically into the three different
  flavors. We also indicate the lifetime of the hidden gaugino that gives the
  best fit to the PAMELA data, assuming an energy density of $\rho_X =
  10^{-3}\rho_\text{DM}$.}
  \label{tab:DHmSUGRA}
\end{table}

As apparent from Tab.~\ref{tab:DHmSUGRA}, the decay into charged
lepton/slepton pairs is dominant in the whole mass range
$M_X\simeq600-1600\GeV$ that we consider. The decay into quarks is suppressed
by the large squark masses, $m_{\tilde{q}}\gtrsim 1.1\TeV$, whereas decay into
$h^0$, $Z^0$ and $W^\pm$ bosons is mainly suppressed by the small mixing
between higgsinos and the hidden gaugino. However, this mixing can become
enhanced when the masses of the higgsinos become comparable to the mass of the
hidden gaugino, which happens around $M_X\sim870\GeV$.

\begin{figure}[h]
  \centering
    \includegraphics[width=\linewidth]{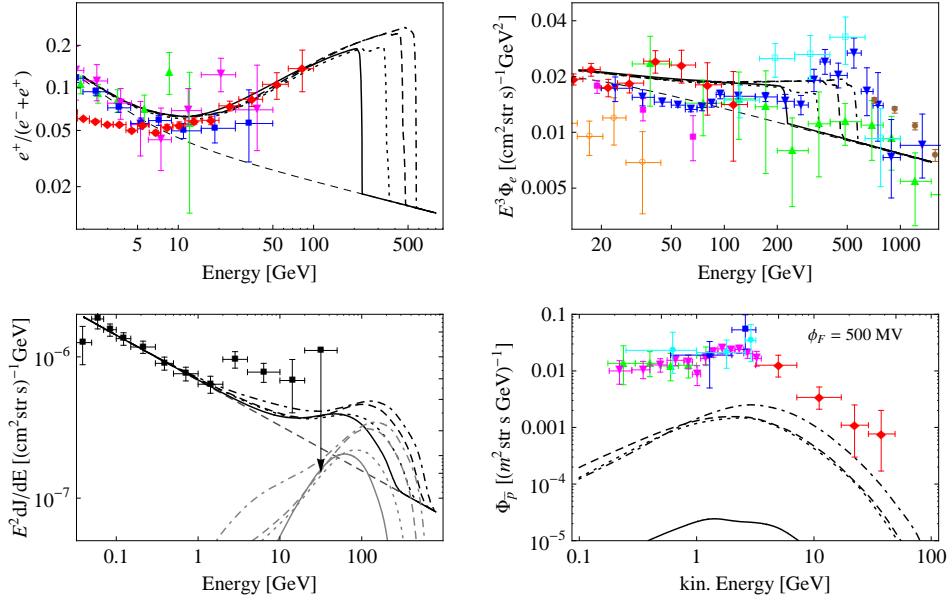}
  \caption{Positron fraction, extragalactic gamma-ray flux, antiproton flux
  and total electron + positron flux from the decay of a hidden gaugino as
  predicted by our mSUGRA scenario. The branching ratios are shown in
  Tab.~\ref{tab:DHmSUGRA}.  The mass of the hidden gaugino varies between
  $600\GeV$ (solid), $800\GeV$ (dotted), $1000\GeV$ (dashed) and $1200\GeV$
  (dot-dashed).}
  \label{fig:DHmSUGRA}
\end{figure}

Our results for the cosmic-ray fluxes are shown in Fig.~\ref{fig:DHmSUGRA} for
hidden gaugino masses between $600\GeV$ and $1200\GeV$, where we adjusted the
lifetime of the hidden gaugino to fit the PAMELA data. For all considered
masses of the hidden gaugino the predictions for the positron fraction are in
qualitatively good agreement with the PAMELA data. At the same time, the
contribution to the antiproton flux lies well below the measurements and hence
is safe in all cases. Furthermore, we obtain contributions to the
extragalactic gamma-ray flux, which are mainly due to $\tau$ decays. They are
compatible with the EGRET measurements but could show up in future experiments
as an excess above background. The total electron+positron flux is also
compatible with the different measurements and we predict a sharp step at high
energies.

\paragraph{Multi-peak structures from cascade decays.}
As already evident in Fig.~\ref{fig:DHmSUGRA}, the energy distribution of
particles produced in cascade decays in general features several peaks. Their
exact position carries information about the masses of the different
intermediate particles.  It is intriguing to speculate that the apparent
double-peak structure of the ATIC data originates from cascade-decaying
particles~\cite{ADD+08}.

In the case of the decaying hidden gaugino, the energy spectrum of positrons
in general possesses two pronounced peaks. These peaks stem from decays with
intermediate selectrons. To simplify the discussion, we will neglect decay
modes that produce $h^0$, $Z^0$ and $W^\pm$ bosons or left-handed sleptons,
and we will assume democratic decay into all three flavors. If we furthermore
assume approximate mass degeneracy for the three right-handed sleptons, we are
left with only three free parameters: the mass of the hidden gaugino $M_X$,
the mass of the lightest neutralino $M_{\chi^0_1}$, and the mass scale of the
right-handed sleptons $M_{\tilde{l}_R}$.

\begin{figure}[ht]
  \centering
    \includegraphics[width=0.6\linewidth]{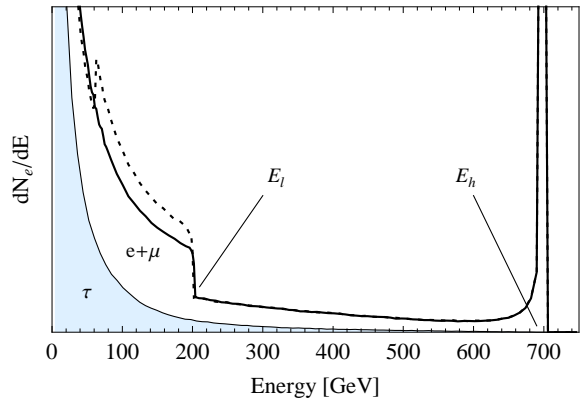}
  \caption{Energy spectrum of positrons from an idealized cascade-decaying
  hidden gaugino. Only two-body decay into right-handed slepton/lepton pairs
  is taken into account. The slepton subsequently decays into the lightest
  neutralino. The spectrum exhibits two pronounced peaks, which we denote by
  $E_h$ and $E_l$. We show plots for a lightest neutralino with mass $150\GeV$
  (solid) and with $1\TeV$ (dashed). The position of the peaks is fixed to
  $E_h=700\GeV$ and $E_l=200\GeV$, as suggested by the ATIC data. The masses
  of the right-handed sleptons follow then from Eq.~\eqref{eqn:NLSPmass}.  We
  also indicate the part of the positrons that comes solely from the tau/stau
  decay channel (blue area).}
  \label{fig:DHspectrum}
\end{figure}

In Fig.~\ref{fig:DHspectrum} we show the corresponding energy spectrum of
positrons for two different sets of particle masses. The two pronounced peaks
are denoted by $E_h$ and $E_l$. Fixing the neutralino mass and the position of
the peaks determines the slepton and hidden gaugino masses according to
\begin{eqnarray}
  M_{\tilde{l}_R}^2 &=& M_{\chi^0_1}^2 + 2 E_l^2
  \left(\sqrt{\left( \frac{E_h}{E_l}-1 \right)^2
  + \left( \frac{M_{\chi^0_1}}{E_l} \right)^2 } - \frac{E_h}{E_l}+1
  \right)\;,
  \label{eqn:NLSPmass}
  \\\nonumber
  M_X &=& E_h + \sqrt{E_h^2+M_{\tilde{l}_R}^2 }\;.
\end{eqnarray}

As a simple attempt to fit the ATIC data with an idealized cascade-decaying
hidden gaugino, we take the values $E_h=700\GeV$ and $E_l=200 \GeV$. After
this, a neutralino mass of $M_{\chi^0_1}=150\GeV$ ($1000\GeV$) implies a
slepton mass of $M_{\tilde{l}_R}=177\GeV$ ($1117\GeV$) and a hidden gaugino
mass of $M_X=1422\GeV$ ($2018\GeV$). 

\begin{figure}[h]
  \centering
  \includegraphics[width=\linewidth]{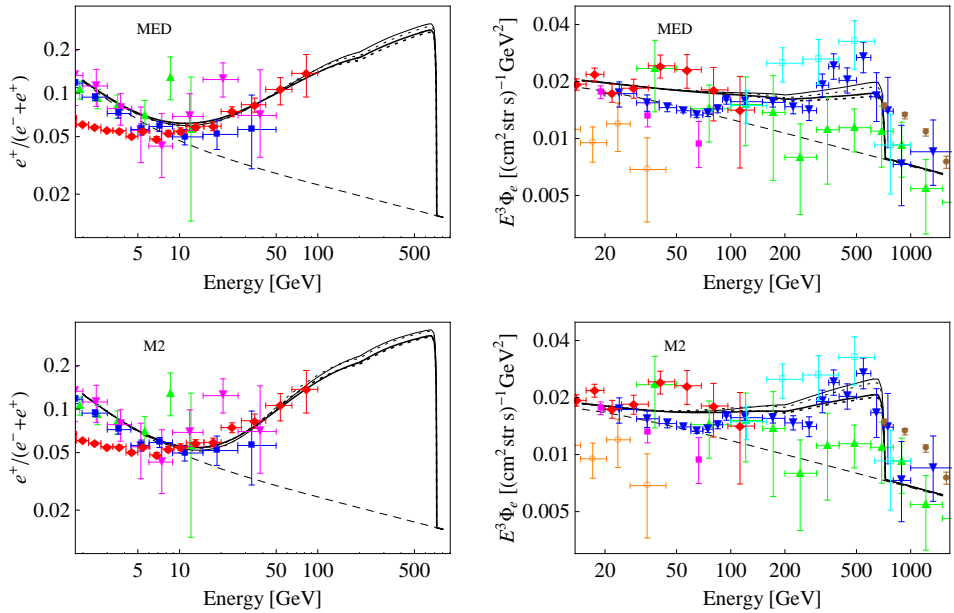}
  \caption{Positron fraction and total electron+positron flux for an idealized
  cascade-decaying hidden gaugino. Like in Fig.~\ref{fig:DHspectrum}, only
  leptonic decay modes are taken into account. We assume democratic decay into
  the three right-handed charged sleptons (thick lines), or into only the
  selectron and smuon (thin lines). The mass of the lightest neutralino varies
  between $150\GeV$ (solid) and $1\TeV$ (dotted). We show the plots for the
  propagation models MED (upper plots) and M2 (lower plots) of
  Ref.~\cite{DLD+08} (see Tab.~\ref{tab:param-positron}).}
  \label{fig:DHbumps}
\end{figure}

The resulting cosmic-ray fluxes for the two neutralino masses are shown in
Fig.~\ref{fig:DHbumps} (upper plots, thick lines), where we used the MED
propagation model. Interestingly, the predicted fluxes are practically the
same for the two cases, although the bump structure is slightly enhanced in
case of the heavier neutralino. The lower plots are based on the M2
propagation model as an exemplary model with a thin diffusion zone, $L=1\kpc$
(as opposed to $L=4\kpc$ in case of the MED model). Since in these scenarios
electrons and positrons are more likely to escape the diffusion zone before
having lost much of their energy, the observable spectrum becomes steeper.
This effect can improve the agreement with the sharp features of the ATIC
data. However, note that propagation models with thin diffusion zones have
problems with the correct prediction of the flux of unstable isotopes like
\fex $^{10}$Be or $^{14}$C~\cite{SMP07}.\footnote{Furthermore, the change of
the propagation model in principle also changes the predictions for the
background of secondary positrons, which were calculated for the above M1, M2
and MED model in Ref.~\cite{D08}. However, in Fig.~\ref{fig:DHbumps} we used
the background from Ref.~\cite{MS98} since the backgrounds obtained in
Ref.~\cite{D08} are meant to be extreme cases.}

Up to now we have assumed a vanishing stau mixing angle and democratic decay
into all three flavors. However, if the stau mixing angle is large the decay
mode into tau/stau pairs would be suppressed, as discussed above. In
Fig.~\ref{fig:DHspectrum} we indicated the part of the positrons that comes
from the stau/tau channel in the case of democratic decay (blue region). A
suppression of this channel can lead to a relative enhancement of the two peak
structure of the cascade decay. This effect is shown by the thin lines in
Fig.~\ref{fig:DHbumps}, where we only took into account the decay modes into
muon/smuon and electron/selectron pairs. As expected, the peaks at high energy
become more pronounced, and the spectrum becomes harder at low energies,
although the effect is not dramatic.

\section{Conclusions}
In this work we have shown that a simple extension of the MSSM by an
additional hidden abelian gauge group $U(1)_X$, which kinetically mixes with
the hypercharge $U(1)_Y$, can account for the observed PAMELA excess if the
kinetic mixing parameter is in the range of $\chi\sim 10^{-(20\dots24)}$. We
also shortly discussed possible origins of such a tiny mixing in scenarios
with warped extra dimensions. Depending on the masses, either the visible
sector neutralino or the hidden gaugino becomes unstable and subject to decay.
We have demonstrated that this decay is dominated by leptonic modes in certain
parameter regions of the MSSM where the sleptons are light (see
Tab.~\ref{tab:DNmSUGRA} and Tab.~\ref{tab:DHmSUGRA}). We found that a decaying
hidden gaugino with a mass around $600-1200\GeV$ can naturally explain the
observed excess in the positron fraction without overproducing antiprotons
(see Fig.~\ref{fig:DHmSUGRA}). Our considerations suggest a preference for
supersymmetric models with relatively light sleptons, as \fex realized in
mSUGRA models which lie in the coannihilation region. In any case, we predict
a contribution to the extragalactic gamma-ray flux, which mainly stems from
tau decays, and which should be observable in future experiments like the
Fermi Gamma-ray Space Telescope.  We also demonstrated that it is difficult to
accommodate the sharp double-peak structure in the ATIC data within our model
and with standard propagation models.

\section*{Acknowledgments}
We would like to thank A. Strong for providing a convenient compilation of
cosmic-ray data. Furthermore, CW gratefully acknowledges G. Sigl for helpful
discussions and the Technische Universit\"at M\"unchen for kind hospitality.
AI and DT would like to thank the Yukawa Institute for Theoretical Physics and
the CERN Theory Division for hospitality during the last stages of this work.
The work of AI and DT was partially supported by the DFG cluster of excellence
``Origin and Structure of the Universe''.\\

During the last stages of our work a preprint appeared that also examines
hidden gauginos in the context of the PAMELA/ATIC anomalies, coming to similar
conclusions~\cite{STY09}.

\frenchspacing
\bibliographystyle{h-physrev3}
\bibliography{}
\end{document}